# Deep Text Classification Can be Fooled

**Bin Liang, Hongcheng Li, Miaoqiang Su, Pan Bian, Xirong Li** and **Wenchang Shi**
School of Information, Renmin University of China, Beijing, China
Key laboratory of Data Engineering and Knowledge Engineering, MOE, Beijing, China
{liangb, owenlee, sumiaoqiang, bianpan, xirong, wenchang}@ruc.edu.cn

## Abstract

In this paper, we present an effective method to craft text adversarial samples, revealing one important yet underestimated fact that DNN-based text classifiers are also prone to adversarial sample attack. Specifically, confronted with different adversarial scenarios, the text items that are important for classification are identified by computing the cost gradients of the input (*white-box* attack) or generating a series of occluded test samples (*black-box* attack). Based on these items, we design three perturbation strategies, namely *insertion*, *modification*, and *removal*, to generate adversarial samples. The experiment results show that the adversarial samples generated by our method can successfully fool both state-of-the-art *character-level* and *word-level* DNN-based text classifiers. The adversarial samples can be perturbed to any desirable classes without compromising their utilities. At the same time, the introduced perturbation is difficult to be perceived.

## 1 Introduction

To improve the robustness of deep neural networks (DNNs), many recent studies [Goodfellow *et al.*, 2015; Kereliuk *et al.*, 2015; Moosavi-Dezfooli *et al.*, 2016; Carlini and Wagner, 2017] have focused on *adversarial samples*, which are well-crafted to cause a trained model to misclassify. For example, Goodfellow *et al.* [2015] showed that a panda image, added with imperceptible perturbations, would be misclassified as a gibbon by GoogLeNet [Szegedy *et al.*, 2015]. However, all existing studies in this field are targeted at DNNs for image or audio classification; the DNNs for natural language processing are seriously underestimated. In fact, text processing plays an important role in modern information analysis. For instance, many phishing webpage and spam detection systems are primarily based on text classification [Whittaker *et al.*, 2010; Slawski, 2014]. A question arises naturally as *whether the text classification DNNs can also be attacked as already done to image or audio classification DNNs*.

Besides fooling the target DNN, we argue that an effective text adversarial sample should meet another two requirements, imperceptible perturbations (i.e., the crafted text can not draw human observers' attention) and utility-preserving. Utility-preserving means that the semantics of the text should remain unchanged and human observers can correctly classify it without many efforts. Consider for instance a spam message advertising something. Its adversarial version should not only fool a spam filter, but also effectively deliver the advertisement. However, satisfying these requirements is highly nontrivial. In essence, text is a kind of *discrete* data, while image or audio data is continuous. Continuous data is tolerant of perturbations to some extent [Goodfellow *et al.*, 2015], while text is not. Even tiny perturbations will turn a character or a word to a completely different one and a sentence to be beyond recognition. As a result, when directly adopting existing multimedia data targeted perturbation algorithms to a text, the resulting text sample may lose its original meaning or even become meaningless for human observers (illustrated in Section 3.1).

In this paper, we present an effective method for crafting adversarial samples against DNN-based text classifiers. Instead of simply overlapping the perturbation and the original input, we design three perturbation strategies (i.e., *insertion*, *modification*, and *removal*) and introduce the *natural language watermarking* technique to elaborately dress up a given text to generate an adversarial sample. In theory, crafting an effective adversarial sample heavily depends on the amount of knowledge about the target classification model. We conduct either *white-box* or *black-box* attack to different adversarial scenarios, to obtain desirable exploitable knowledge. For the model whose implementation can be freely and completely analyzed by the adversaries, the cost gradients of the input are computed to effectively determine what, where and how to insert, modify or remove. Otherwise, if the target model cannot be directly analyzed, we generate some occluded test samples to probe it and obtain the above exploitable knowledge.

Without loss of generality, we pick two representative text classification DNNs, i.e., a character-level model [Zhang *et al.*, 2015] and a word-level one [Kim, 2014], as attack targets. The attack experiments show that despite the conciseness, our method can perform effective source/target misclassification attack against both DNNs and the adversarial samples generated by our three strategies satisfy all the requirements, i.e., fooling the target DNN, imperceptible perturbations and utility-preserving. Our work effectively reveals that

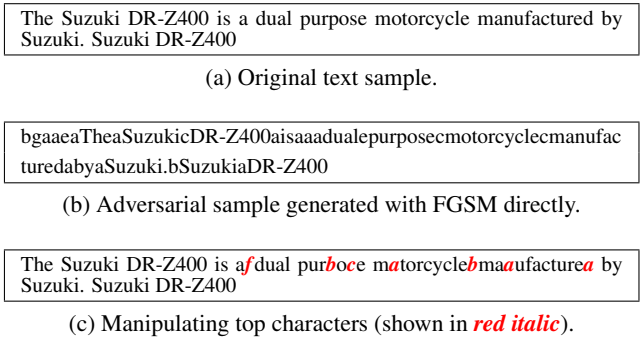

The Suzuki DR-Z400 is a dual purpose motorcycle manufactured by Suzuki. Suzuki DR-Z400

(a) Original text sample.

bgaaeaTheaSuzukicDR-Z400aisaaadualepurposecmotorcyclecmanufacturedabyaSuzuki.bSuzukiaDR-Z400

(b) Adversarial sample generated with FGSM directly.

The Suzuki DR-Z400 is a*f* dual pur*bo*ce m*a*torcycle*b*ma*a*ufacture*a* by Suzuki. Suzuki DR-Z400

(c) Manipulating top characters (shown in **red italic**).

Figure 1: Adversarial text samples generated with FGSM.

DNN-based text classifiers are definitely vulnerable to the adversarial sample attack.

## 2 Target Models and Datasets

Two representative DNN models and some corresponding datasets are chosen as the experiment targets to evaluate the effectiveness of the proposed method. One is a character-level model [Zhang *et al.*, 2015] and the other is word-level [Kim, 2014]. The character-level DNN is trained on a DBpedia ontology dataset, which contains 560,000 training samples and 70,000 testing samples of 14 high-level classes, such as *Company*, *Building*, *Film* and so on. Before feeding an input text into the network, every character of the text is quantized as a vector using one-hot representation. For example, the character 'c' is encoded as the vector (0, 0, 1, 0, ..., 0), in which only the third dimension is set to one. Through six convolutional layers and three fully-connected layers, the input is finally projected into a vector indicating the classification confidences of the 14 classes.

The word-level model consists of one convolutional layer, followed by a max pooling layer and a fully connected layer with dropout, and last a softmax output layer. Before being fed to the model, all the words within the input text will be embedded to some 300-dimensional pre-trained vectors [Mikolov *et al.*, 2013]. The model is tested on several datasets, including MR, CR and MPQA. The MR dataset is a movie review repository (containing 10,662 reviews) while CR contains 3,775 reviews about products, e.g. a music player. Reviews from both datasets can be categorized as either *Positive* or *Negative*. MPQA contains 10,606 opinions, and each of them is labeled as *Objective* or *Subjective*.

## 3 White-box Attack

### 3.1 Overview

Goodfellow *et al.* [2015] first presented an effective gradient-based method, called *fast gradient sign method* (FGSM for short), to craft adversarial image samples. In the FGSM algorithm, the introduced perturbation $\Delta x$ is computed as the sign of the model's cost gradients, i.e., $\varepsilon sign(\nabla_x J(F, x, c))$, where $c$ is the true class of input $x$, $J(F, x, c)$ is the cost function used to train the DNN $F$ and $\varepsilon$ is set to be small enough to make $\Delta x$ undetectable. Simply overlapping $\Delta x$ and $x$ regardless of gradients' magnitude has been proven effective to craft adversarial samples. Sometimes, only manipulating the input pixels with large gradient magnitude can also do the trick since a pixel with larger gradient magnitude often contributes more to the current prediction, as illustrated in [Liang *et al.*, 2017].

| Rank | HTP | Freq. | Rank | HTP | Freq. |
|------|------|-------|------|------|-------|
| 1 | historic | 7279 | 6 | house | 2447 |
| 2 | building | 4954 | 7 | built | 1927 |
| 3 | church | 3978 | 8 | is a | 1761 |
| 4 | Register | 3418 | 9 | museum | 1239 |
| 5 | located | 2604 | 10 | is | 1101 |

Table 1: Top ten HTPs of *Building* class.

We also leverage the cost gradient to craft adversarial samples, but in the text domain. However, the technique is not directly applicable for text classification. As illustrated in Figure 1(a), the original text is a short description of a motorcycle and correctly classified as *Means of Transportation* by the DNN model [Zhang *et al.*, 2015]. Although the prediction can be altered to *Company* by FGSM, the new text, see Figure 1(b), is unreadable. One might consider manipulating only a few characters with the highest gradient magnitude. Still, the generated text is unnatural with noticeable perturbations (see Figure 1(c)). It is clear that more sophisticated strategies are needed to craft adversarial samples for the text data. Specifically, in the white-box scenario, we first identify text items that are important for classification according to their cost gradients, and then leverage these items, along with natural language watermarking technique, to decide what to insert, modify or remove, where to insert and how to modify.

### 3.2 Identifying Classification-important Items

The foundation of our white-box attack lies in identifying the text items that possess significant contribution to the classification by leveraging the cost gradient. Take the character-level DNN for example. First, the backpropagation algorithm is employed to compute the cost gradients $\nabla_x J(F, x, c)$ for every training sample $x$. Consequently, we get the cost gradients of every dimension in all character vectors of these samples. Then we term characters containing the dimensions with maximum highest magnitude *hot characters*. Per sample we empirically select the top 50 hot characters. Subsequently, hot words containing more than or equal to three hot characters can be identified through a simple scan and any adjacent words will be assembled as a hot phrase if they are all hot words. Hot words that are not assembled will be treated as hot phrases, too. Finally, for all training samples, the hot phrases obtained previously will be collected together according to samples' labels. From them, we determine the most frequent phrases, called *Hot Training Phrases* (HTPs). Take the *Building* class for example, its top ten HTPs are illustrated in Table 1. HTPs for the word-level DNN can be obtained in a similar way, only this time hot phrases are directly identified by looking for the word vectors possessing the maximum highest gradient magnitudes.

HTPs shed light on what to insert, but where to insert, remove and modify remains unclear. Fortunately, obtaining

> The Uganda Securities Exchange (USE) is the ***historic*** principal stock exchange of Uganda. It was founded in June 1997. The USE is operated under the jurisdiction of Uganda's Capital Markets Authority which in turn reports to the Bank of Uganda, Uganda's central bank. The exchange's doors opened to trading in January 1998. At the time, the exchange had just one listing, a bond issued by the East African Development Bank. Uganda Securities Exchange

Figure 2: An adversarial text sample generated by inserting just one HTP (99.7% *Company* to 88.6% *Building*).

> Hot Issue is the second Korean EP by South Korean boy band Big Bang, released under YG Entertainment***, an entertainment company founded in 1996 in Seoul, South Korea***. Big Bang's first EP Always further established the group's popularity in South Korea, with the single Last Farewell topping online charts for 8 consecutive weeks, sold over 5 million digital downloads. The group's leader, the then 20-year-old G-Dragon produced and wrote the lyrics for all tracks on Hot Issue. The song is a blend of trance hip-hop beats and pop melodies. Hot Issue (EP)

Figure 3: An adversarial text sample generated by inserting a parenthesis based on a fact (99.9% *Album* to 94.0% *Company*).

> The APM 20 Lionceau is a two-seat very light aircraft manufactured by the French manufacturer Issoire Aviation. Despite its classic appearance it is entirely built from composite materials especially carbon fibers. Designed by Philippe Moniot and certified in 1999 (see EASA CS-VLA) this very light (400 kg empty 634 kg loaded) and economical (80 PS engine) aircraft is primarily intended to be used to learn to fly but also to travel with a relatively high cruise speed (113 knots). ***Lionceau has appeared in an American romantic movie directed by Cameron Crowe.*** A three-seat version the APM 30 Lion was present-ed at the 2005 Paris Air Show. Issoire APM 20 Lionceau such information is similar to identifying HTPs. Given a text sample, we still employ the backpropagation algorithm to locate hot phrases with significant contribution to the current classification, and those phrases are recognized as *Hot Sample Phrases* (HSPs). HSPs imply where to manipulate to craft an effectual adversarial sample.

Figure 4: An adversarial text sample generated by inserting a forged fact (99.9% *Means of Transportation* to 90.2% *Film*).

### 3.3 Attacking Character-level DNN

Based on HTPs and HSPs, three perturbation strategies, i.e., *insertion*, *modification* and *removal*, are used alone or in combination to craft an adversarial sample for a given text. It is worth pointing out that *the proposed method allows us to arbitrarily choose a class of interest as the target class to launch a source/target misclassification attack* rather than just altering the output classification to any class different from the original, as presented in [Goodfellow *et al.*, 2015].

**Insertion Strategy**

For a given text $t$, if $F(t) = c$, the goal of the strategy is to insert some new text items (attack payload) in $t$, which can effectually upgrade the classification confidence of $t$ for the class $c'$ of interest ($c' \neq c$) and hopefully downgrade the confidence of the original class $c$ accordingly. HTPs are used as the key elements to construct our attack payload. As Figure 2 shows, inserting just one HTP ("historic") just before the HSP ("principal stock exchange of Uganda. It was founded") can successfully let a text describing a company be misclassified as the *Building* class with 88.6% confidence.

In practice, multiple insertions are often needed. However, directly inserting multiple HTPs into a text is likely to hurt its utility and readability. To address the problem as well as enrich the means of attacking, we introduce the NL (Natural Language) watermarking technique. The technique can stealthily embed an ownership watermark into a plain text via manipulating it semantically or syntactically, such as replacing words with their synonyms or typos [Topkara *et al.*, 2006; 2007], paraphrasing representation [Barzilay and McKeown, 2001], adding presuppositions [Vybornova and Macq, 2007], and inserting semantically empty phrases [Atallah *et al.*, 2001]. Though the goal of our attack is essentially different from that of NL watermarking, we can borrow its idea to craft adversarial samples. In fact, the perturbation can be regarded as a kind of watermark and embedded into the sample in similar ways.

Here, we extend the ideas of inserting presuppositions [Vybornova and Macq, 2007] and semantically empty phrases [Atallah *et al.*, 2001] to perturb the target text sample. Presupposition is an implicit information that can be considered well-known to the readers, and a semantically empty phrase is a dispensable component. With or without them, the text's meaning remains unchanged. In general, we consider to introduce multiple HTPs by assembling them into a syntax unit and inserting it in a proper position. The new unit can be crafted as a dispensable fact (see Figure 3) or even a *forged fact* (see Figure 4) that does not hurt the text's primary semantics.

Specifically, by searching Internet or some databases of facts, e.g., [Suchanek *et al.*, 2007], we can get some facts that are closely related to the insertion point and embody some desirable HTPs of the target class as well. For example, as shown in Figure 3, we google the company name "YG Entertainment" along with some combinations of HTPs and a related fact that carries three HTPs of the *Company* class ("company", "founded" and "entertainment") can be easily found from Wikipedia. Inserting it after the company name can craft an effectual adversarial sample, without drawing human observer's attention. When a proper fact is not available, we present a new concept, called *forged fact*, to wrap desirable HTPs. The forged fact can be created by reforming some real things related to the HTPs to make people believe it really happened. Furthermore, we exclude forged facts that can be disproved by retrieving their opposing evidences on Internet. Figure 4 presents a forged fact carrying desirable HTPs ("romantic", "movie", "directed by" and "American") that fools the target DNN.

**Modification Strategy**

The modification strategy is to affect the model output by slightly manipulating some HSPs in the input. In theory, the modification should increase the cost function $J(F, t, c)$ and meanwhile decrease $J(F, t, c')$. In other words, the modification should follow the direction of the cost gradient $J(F, t, c)$, yet against the direction of $J(F, t, c')$.

In order to perform the modification without getting human observer's attention, we adopt the typo-based watermarking technique [Topkara *et al.*, 2007]. Specifically, an HSP is modified in two ways: (1) replaced with its common mis-

> Maisie is a comedy f*li*m property MGM originally purchased for Jean Harlow but before a shooting script could be completed Harlow died in 1937. It was put on hold until 1939 when Ann Sothern was hired to star in the project with Robert Young as leading man. It is based on the novel Dark Dame by Wilson Collison. It was the first of ten films starring Sothern as Maisie Ravier. In Mary C. Maisie (film) spellings, and (2) some characters of it be changed to ones in similar visual appearance, e.g., the lower-case letter 'l' (el) be replaced with the digit '1' (one). As shown in Figure 5, typo-based modifications can cause dramatic prediction errors. The phrase "comedy film property" is an HSP of the original class (*Film*) and the typo "flim" is obtained from a misspelling corpus [Mitton, 1985]. Furthermore, as illustrated in Table 2, replacing "film" with "flim" does follow the desirable gradient directions.

Figure 5: An adversarial text sample generated by introducing a common misspelling (99.6% *Film* to 99.0% *Company*).

| Modification "film" → "flim" | | Cost Gradient | |
|---|---|---|---|
| | | Source Class (*Film*) | Target Class (*Company*) |
| 'i' → 'l':(..., 1, ..., 0, ...) | 1 → 0 ↓ | -0.01913 ↓ | 4.61543 ↑ |
| → (..., 0, ..., 1, ...) | 0 → 1 ↑ | 0.00707 ↑ | -1.71717 ↓ |
| 'l' → 'i':(..., 0, ..., 1, ...) | 0 → 1 ↑ | 0.01457 ↑ | -3.50519 ↓ |
| → (..., 1, ..., 0, ...) | 1 → 0 ↓ | -0.03177 ↓ | 7.62243 ↑ |

Table 2: The Direction of the Cost Gradients.

### Removal Strategy
The removal strategy alone may not be effective enough to alter the prediction, but can largely downgrade the confidence of the original class. Since arbitrarily eliminating HSPs from the input often compromises its meaning, only words of the HSPs that play as a supplementary role, e.g., an inessential adjective or adverb, can be removed. As Figure 6 shows, removing "British" (part of the HSP "seven-part British television series") can result in a confidence decline of 35.0%.

### Combination of Three Strategies
As illustrated in Figure 6, changing the output classification by the removal strategy alone is often difficult. However, by combining it with the other strategies, excessive modifications or insertions to the original text can be avoided. In practice, we often combine the above three strategies to craft subtle adversarial samples. Take Figure 7 for example, by removing an HSP, inserting a forged fact and modifying an HSP, the output classification can be successfully changed, but applying any above perturbation alone fails. Specifically, the removal, insertion and modification only downgrade the confidence by 27.3%, 17.5% and 10.1% respectively, keeping the prediction class unchanged.

### 3.4 Attacking Word-level DNN
Attacking the word-level DNN is similar to attacking the character-level one. As shown in Figure 8, leveraging the obtained HTPs and HSPs, the three perturbation strategies are proven to be effective. Figure 8(a) presents a positive

> Edward & Mrs. Simpson is a seven-part ***British*** television series that dramatises the events leading to the 1936 abdication of King Edward VIII of the United Kingdom who gave up his throne to marry the twice-divorced American Wallis Simpson. The series made by Thames Television for ITV was originally broadcast in 1978. Edward Fox played Edward and Cynthia Harris portrayed Mrs. Simpson. Edward & Mrs. Simpson

Figure 6: Lower the confidence of the original class by removing a word from an HSP (95.5% *Film* to 60.5% *Film*).

> The Old Harbor Reservation Parkways are three ***historic*** roads in the Old Harbor area of Boston. ***Some exhibitions of Navy aircrafts were held here.*** They are part of the Boston parkway system designed by Frederick Law Olmsted. They include all of William J. Day Boulevard running from Cast*I*e Island to Kosciuszko Circle along Pleasure Bay and the Old Harbor shore. The part of Columbia Road from its northeastern end at Farragut Road west to Pacuska Circle (formerly called Preble Circle). Old Harbor Reservation movie review from MR dataset. Inserting one HTP "boring" from the *Negative* class successfully alters the prediction. As demonstrated in Figure 3 and 4, single insertion may not always be sufficient to craft an adversarial sample, and in such cases, a proper parenthesis that wraps several desired HTPs is needed. In Figure 8(b), we compose a clause with two *Positive* HTPs, i.e., "awesome" and "amazing", and insert it into a negative review of a music player (from CR). The resultant text is misclassified as *Positive* with 87.3% confidence. Sometimes, the modification strategy can also be performed by paraphrasing representations, as done in NL watermarking [Barzilay and McKeown, 2001]. Taking Figure 8(c) for example, by simply paraphrasing the phrase "different from" to "not" (a *Negative* HTP), we make a positive customer review be misclassified as *Negative*. As for the removal strategy, a sample (Figure 8(d)) from MPQA dataset shows that its adversarial sample can be crafted by simply removing a phrase which contains an *Objective* HSP "promotion".

Figure 7: Combination of three strategies (83.7% *Building* to 95.7% *Means of Transportation*).

## 4 Black-box Attack
In the black-box scenario, internal knowledge of the target model (e.g., cost gradients) is no longer available. In such cases, we borrow the idea of the *fuzzing* technique [Sutton *et al.*, 2007] to implement a blind test to locate HTPs and HSPs. By sophisticatedly generating a number of malformed inputs, fuzzing can trigger unexpected system behaviors (e.g., a system crash) to find potential security bugs, even without knowing the detailed knowledge about the target system.

Similarly, in our proposed method, some test samples are purposefully generated to probe the target model. Since in the black-box scenario, we don't know the technique details about the model, including the input preprocessing, the test sample generation policy should remain the same no matter the model is character-level or word-level. In practice, we employ one straightforward generation policy to avoid producing too many test samples and being blocked in certain cases. As shown in Figure 9, to generate test samples for a given seed sample, we occlude its words one by one with a sequence of *whitespaces*, whose length is exactly the same

> spider-man is better than any summer blockbuster we had to endure last ***boring*** summer , and hopefully , sets the tone for a summer of good stuff . if you're a comic fan , you can't miss it . if you're not , you'll still have a good time .

(a) Inserting one HTP (97.8% *Positive* to 90.2% *Negative*).

> no in box accessories – an arm strap , belt strap , case , or anything would be nice , but nope . . . moveable parts – this makes me almost want to keep my 3rd gen over a 4th ; . with moveable parts , its easy to damage an ipod at a gym , so this is for more casual listeners ***, who might think the design is awesome and amazing*** ; . note the ipod mini doesn 't have moveable parts and is made of titanium .

(b) Inserting multiple HTPs (99.7% *Negative* to 87.3% *Positive*).

> the battery life last es 12 hours , and charges quickly . ( as mentioned above , the initial charge is a minimum of only two hours - not too long at all ! . ) once you 've downloaded your music , you can access the tracks by artist , genre , album , composer ( this comes in handy when the actual composer is ***different from not*** the artist - e . g . covers of songs ) , etc .

(c) Paraphrasing a phrase (83.9% *Positive* to 92.0% *Negative*).

> ~~*promotion of world security ,*~~ improvement of economic conditions , defusing regional and global crises and conflicts , checking unleashed competition for rearmament , reduction of the gap between the mainstream and peripheral countries

(d) Removing a phrase (87.8% *Objective* to 69.3% *Subjective*).

Figure 8: Attacking the word-level DNN.

| No. | Samples | Conf. |
|---|---|---|
| seed | Edward & Mrs. Simpson is a seven-part ... | 95.5% |
| 1 | ⊔⊔⊔⊔⊔⊔ & Mrs. Simpson is a seven-part ... | 98.2% |
| 2 | Edward ⊔ Mrs. Simpson is a seven-part ... | 97.8% |
| ... | ... | ... |
| 8 | ... seven-part ⊔⊔⊔⊔⊔⊔⊔ television series ... | 68.6% |
| 9 | ... seven-part British ⊔⊔⊔⊔⊔⊔⊔⊔⊔ series ... | 53.3% |
| ... | ... | ... |
| 70 | ... Mrs. Simpson. Edward & Mrs. ⊔⊔⊔⊔⊔⊔⊔ | 97.8% |

Table 3: The test samples generated from a seed.

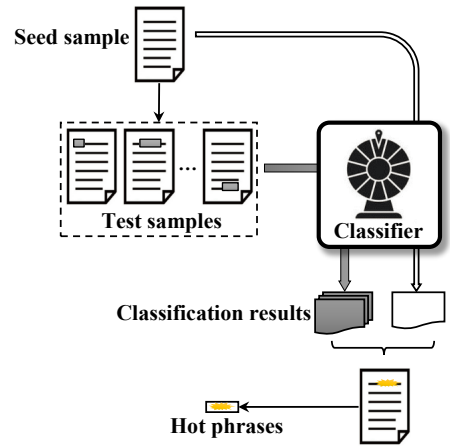

Figure 9: Identifying hot phrases with the black-box test.

| Rank | *Positive* Review | | *Negative* Review | |
|---|---|---|---|---|
| | White-box | Black-box | White-box | Black-box |
| 1 | great | great | not | not |
| 2 | easy | easy | problem | 't |
| 3 | good | good | no | problem |
| 4 | excellent | excellent | problems | little |
| 5 | best | and | slow | no |
| 6 | nice | best | too | bit |
| 7 | not | love | annoying | annoying |
| 8 | love | nice | flaw | flaw |
| 9 | awesome | no | useless | problems |
| 10 | pleased | well | little | slow |

Table 4: Top ten HTPs of CR dataset.

as the occluded word. Choosing whitespaces for occlusion is because in many natural languages, e.g. English, redundant whitespaces often contribute nearly nothing to the semantics of the texts, and the "same length" restriction is to preserve the structural integrity of the whole text as far as possible. In some DNN-based text classifiers, e.g., [Zhang *et al.*, 2015], the classification of a text is impressionable to its structure. Hence, using only one whitespace for occlusion or directly removing a word may cause unwanted problems. We feed all the test samples to the target model and record their classification results. By comparing the classification result of a test sample with the seed, we can learn how much deviation an occluded word can cause. The larger the deviation is, the more importance the corresponding word attaches to the correct classification. The words that can bring largest deviations are identified as HSPs for the seed sample.

Taking the 70-word text in Figure 6 as a seed, 70 test samples can be generated with the proposed scheme, as Table 3 shows. After feeding them to the target model [Zhang *et al.*, 2015], we find that occluding "television" and "British" can cause the largest confidence declines (dropping by 42.2% and 26.9% respectively), and the two words can be identified as the HSPs of the text. By using them, we develop the same perturbation as illustrated in Figure 6, i.e., removing "British". HTPs can be identified in a similar way. In our experiments, we apply the occlusion scheme to the training sets and only select one hot word (the word that being occluded can cause the largest deviation) from each sample. All the hot words are collected according to the samples' labels, and the most frequent ones are recognized as HTPs. The experiment results show that the black-box testing can achieve similar performance in identifying HTPs with the white-box way. Table 4 presents the top ten HTPs for the two classes of CR dataset, which are found by white-box and black-box method respectively. We can see that seven out of top ten HTPs are the same (shown in gray background) for *Positive* class and eight for *Negative*.

In fact, all the HTPs and HSPs used to craft the adversarial samples from Figures 2 ∼ 8 can be identified through black-box testing. The same perturbation strategies (i.e. *insertion*, *modification* and *removal*) can be adopted to generate the presented adversarial samples. We believe that black-box attack is as effective as the white-box one.

| No. | Source Class No. | Target Class No. | Inserted HTPs | Modified HSPs | Removed HSPs |
|---|---|---|---|---|---|
| 1 | 6#(99.9%) | 1#(94.4%) | 3 | 0 | 0 |
| 2 | | 2#(76.3%) | 5 | 0 | 1 |
| 3 | | 3#(68.2%) | 5 | 3 | 1 |
| 4 | | 4#(84.7%) | 3 | 3 | 1 |
| 5 | | 5#(82.4%) | 4 | 0 | 1 |
| 6 | | 7#(86.3%) | 3 | 2 | 1 |
| 7 | | 8#(70.0%) | 2 | 3 | 1 |
| 8 | | 9#(89.7%) | 3 | 2 | 1 |
| 9 | | 10#(81.0%) | 3 | 2 | 1 |
| 10 | | 11#(82.4%) | 3 | 0 | 0 |
| 11 | | 12#(66.3%) | 3 | 2 | 1 |
| 12 | | 13#(82.4%) | 4 | 0 | 0 |
| 13 | | 14#(80.2%) | 4 | 2 | 1 |
| 14 | 1#(99.7%) | 7#(88.6%) | 1 | 0 | 0 |
| 15 | 2#(99.3%) | 13#(72.0%) | 2 | 1 | 0 |
| 16 | 3#(99.8%) | 12#(84.2%) | 2 | 1 | 0 |
| 17 | 4#(99.9%) | 3#(77.5%) | 2 | 1 | 1 |
| 18 | 5#(99.8%) | 4#(95.9%) | 3 | 1 | 0 |
| 19 | 7#(83.7%) | 6#(88.7%) | 2 | 1 | 1 |
| 20 | 8#(99.5%) | 7#(96.3%) | 1 | 0 | 0 |
| 21 | 9#(99.9%) | 7#(98.2%) | 1 | 1 | 0 |
| 22 | 10#(84.6%) | 6#(94.2%) | 0 | 0 | 1 |
| 23 | 11#(85.3%) | 8#(81.7%) | 1 | 0 | 1 |
| 24 | 12#(99.9%) | 1#(94.0%) | 2 | 0 | 0 |
| 25 | 13#(99.6%) | 1#(99.0%) | 0 | 1 | 0 |
| 26 | 14#(99.9%) | 13#(87.1%) | 3 | 0 | 1 |
| Avg. | 98.1% | 84.7% | 2.5 | 1.0 | 0.6 |

Table 5: Experiments for Q1 and Q2.

## 5 Evaluation

We evaluate the effectiveness of our method by answering the following questions.

***Q1: Can our method perform effective source/target misclassification attack?*** Among the test sets, the DBpedia ontology dataset is the only multi-class set. We randomly select a sample from it, originally classified as *Means of Transportation* (class 6#), and manage to perturb the sample to the other 13 classes using the three strategies (see Table 5, No. 1∼13). In addition, we randomly select one sample from each of the 13 classes (except *Means of Transportation*) and manage to craft desirable adversarial samples as well (see Table 5, No. 14∼26). Those experiments indicate that our method can perform effective source/target misclassification attack.

***Q2: Can the adversarial samples avoid being distinguished by human observers and still keep the utility?*** We perform a single-blind user study with 23 students from our university as subjects. They have no prior knowledge about this project and each of them is provided with 20 text samples, half of which are with perturbations. The subjects don't know which samples are affected. They are asked to manually classify each sample. Further, if they consider a sample is artificially modified, they are asked to pinpoint where the modification is performed.

For the ten original samples, the subjects classify them with an averaged accuracy of 94.2%, while for the ten affected samples the accuracy is 94.8%. The comparable performance shows that the utility is indeed preserved in the adversarial samples. In total, there are 240 places marked as modified, with 12 successful matches. Recall that our method yields 594 changes. Our modifications can be detected by the human observers with an accuracy of 12/240=5.0% and a recall of 12/594=2.0%. The result suggests that the crafted adversarial samples are difficult to be perceived.

***Q3: Is our method efficient enough?*** The white-box attack took 116 hours in total to compute the cost gradient and identify HTPs for all the 14 classes of the DBpedia dataset (8.29 hours per class) on a desktop computer. For the black-box attack, generating test samples and identifying the HTPs cost 107 hours (7.63 hours per class). For the other relatively small datasets (i.e., MR, CR and MPQA), we can get their HTPs within several minutes. Crafting one adversarial sample takes about 15 minutes. In practice, attackers are willing to spend more time to craft a desirable adversarial sample.

***Q4: White-box and black-box, which is more powerful?*** Our experiments show that HTPs obtained in the two ways are highly overlapped, about 80% of which are the same, and for a given sample, the obtained HSPs are often the same. Furthermore, in practice, the white-box analysis may identify some hot phrases that are missed in the black-box test, and vice versa. For example, two exploitable *Negative* HTPs "disappointed" and "terrible" from the CR dataset are recognized in the white-box way, which are missed in the black-box way. At the same time, two exploitable *Positive* HTPs "beautiful" and "charming" from the MR dataset can only be hit in the black-box test. The four words can all be leveraged to perform successful evasion attacks. These evidences imply that both ways are effective and complementary to each other.

## 6 Related Work

Many studies have being focused on crafting adversarial samples to bypass DNN-based classifiers, and various generation techniques have been proposed, including *gradient*-based [Goodfellow *et al.*, 2015; Kereliuk *et al.*, 2015], *decision function*-based [Moosavi-Dezfooli *et al.*, 2016] and *evolution*-based [Nguyen *et al.*, 2015] approaches. The gradient-based method is most efficient and easy to use. Papernot et al. [2016] introduced a defense named *defensive distillation* to adversarial samples. Two networks were trained as a distillation, where the first network produced probability vectors used to label the original dataset, while the other was trained using the newly labeled dataset. As a result, the effectiveness of adversarial samples can be substantially reduced. However, as pointed in [Carlini and Wagner, 2017], how to construct defenses that are robust to adversarial examples remains open. A feasible solution is to improve the robustness of DNNs by making it harder for attackers to craft adversarial samples rather than eliminating them thoroughly. The adversarial training is a straightforward defense technique which uses as many as possible adversarial samples during the training process as a kind of regularization [Goodfellow *et al.*, 2015; Kereliuk *et al.*, 2015]. This can make it harder for attackers to generate new adversarial samples. All of these studies are targeted at image DNN-based classifiers. Text as discrete data is sensitive to perturbation and the above methods are not directly applicable for text.

Recently, a few researchers began to pay close attention to adversarial samples for text data. Jia and Liang [2017] created adversarial examples for sixteen reading comprehension systems by appending distracting sentences to a given paragraph. The resulting paragraph can cause the target system to output a wrong answer. However, the appended sentence tends to be incompatible with the whole target paragraph. Gong *et al.* [2018] adopted image targeted perturbation algorithms (e.g., FGSM) to disturb the word embedding space of given text samples and reconstructed the space via *nearest neighbor search*. In this way, the generated text can be free from gibberish words. However, according to the authors, their method depends heavily on the size and quality of the embedding space. Besides, from some presented adversarial texts, we can see there are a considerable number of perturbations that make the text unnatural. Gao *et al.* [2018] proposed a black-box attack algorithm to generate small perturbations. In general, the algorithm first determines the important tokens according to the scoring functions they defined, and then applies some character-level transformations (i.e. *swap*, *substitution*, *deletion* and *insertion*) to the identified tokens. Although this kind of character-level attack strategy adopted by this algorithm can successfully generate certain adversarial samples, the introduced perturbations are limited to typos. In practice, introducing excessive typos will heavily harm the utlity of the samples.

## 7 Discussion and Conclusion

In this study, we've revealed one important yet underestimated fact that DNN-based text classification is surprisingly vulnerable to adversarial sample attack, so its robustness should be seriously considered. Specifically, we conduct either white-box or black-box attack to different adversarial scenarios and craft effectual text adversarial samples in both ways. Still, our work can be further refined. Arras *et al.* [2016] leverage a technique called layer-wise relevance propagation (LRP) to locate words that have positive impact as well as words that may bring negative impact for a specific predication of a DNN. We believe such technique can also be adopted for identifying HTPs and HSPs in white-box attack. Our black-box attack assumes access to training set and demands confidence degree for feedback, however, such resources may not always be available. Some studies have demonstrated that adversarial samples can transfer from one model to another, even if the second model has a different architecture or was trained on a different set [Szegedy *et al.*, 2014; Goodfellow *et al.*, 2015], which implies that our black-box attack may can be conducted in an indirect way, e.g., training and fooling a substitute model.

So far, our method requires some human efforts when crafting an adversarial sample. Such amounts of human efforts are acceptable for common attack scenarios. To support adversarial training and investigate the possibility of large-scale attacks, in the future, we will address how to automatically craft adversarial text samples.

## Acknowledgments
The authors would like to thank the anonymous reviewers for their insightful comments on the preliminary version of this paper. The work is supported by National Natural Science Foundation of China (NSFC) under grants 91418206, 61672523, and 61472429, National Science and Technology Major Project of China under grant 2012ZX01039-004.